\documentclass{PoS}
\usepackage{natbib}
\usepackage{epsfig}
\usepackage{amsmath}

\title{H.E.S.S. Observations of strong flaring activity of Mrk421 in February 2010}

\ShortTitle{H.E.S.S. Observations of strong flaring activity of Mrk421 in February 2010}

\author{Martin Tluczykont (for the H.E.S.S. Collaboration)\\%
       Universit\"at Hamburg\\
       E-mail: \email{martin.tluczykont@physik.uni-hamburg.de}}

\abstract{
The high-frequency peaked BL Lac object Mrk421 has shown a strong outburst of activity during February 2010.  This high state, first detected by VERITAS on 16th of February, was followed up by H.E.S.S. during four subsequent nights at an average zenith angle of 62.4$^\circ$.
In the first night of observations by H.E.S.S., the average flux level above 1 TeV was 4 times the level of the Crab Nebula flux, decreasing to a third of this value over the subsequent observation nights.  The H.E.S.S. data shows an energy spectrum compatible with previous measurements, extending up to 20 TeV. 
The observed spectral shape can be well described by a powerlaw with
a spectral index of $\Gamma = 2.05\,\pm\,0.22$,
an exponential cutoff $E_{cut} = (3.4\,\pm\,0.6)$\,TeV
and a flux normalization $\phi_0 = (1.96\,\pm\,0.32)\,10^{-11}$\,cm$^{-2}$s$^{-1}$TeV$^{-1}$
at decorrelation energy $E_0 = 2.739$\,TeV.
The observed integral flux varies on a night-by-night basis between 1.4 and 4.8 times the flux of the
Crab Nebula.
}

\FullConference{25th Texas Symposium on Relativistic Astrophysics \\
                December 6-10, 2010\\
                Heidelberg, Germany}

\begin{document}

\section{Introduction}
The broadband emission of AGN was observed by different experiments reaching from radio waves
to the very high energy (VHE, E$>$100\,GeV) gamma-ray band.
In the unified scheme of AGN \citep[e.g.][]{1984ARA&A..22..471R,1995PASP..107..803U},
the ``central engine'' of these objects consists
of a super-massive black hole (up to 10$^9$\,M$_\odot$) surrounded
by a thin accretion disk and a dust torus.
Two opposed relativistic plasma outflows (jets) perpendicular to the accretion disk have been observed
in radio-loud AGN.
Flux variability of AGN in the VHE gamma-ray regime was measured down to
time-scales of minutes \citep{2008PhLB..668..253M,2007ApJ...664L..71A}.
Variability studies can contribute to the understanding of
the intrinsic acceleration mechanisms of AGN \citep[e.g.][]{2001ApJ...559..187K,2002A&A...393...89A}.

Mrk\,421 (located at a redshift of $z=0.03$) belongs to the class of highly variable BL Lac objects and
was the first active galactic nucleus (AGN) to be detected in the VHE gamma-ray
regime \citep{1992Natur.358..477P}.
Today, monitoring observations of VHE gamma-rays from AGN are
carried out on a regular basis by different imaging air Cherenkov telescope experiments.
Mrk\,421 is the object with the longest lightcurve coverage by VHE experiments, its combined lightcurve
spanning almost two decades, with indications for a log-normal flux variability behaviour
\citep{2010A&A...524A..48T}
which is indicative of an underlying multiplicative acceleration process, such as accretion of matter
onto the central black hole \citep[see][and references therein]{2009A&A...503..797G}.

The measured flux in the VHE band from AGN (or any distant object) is attenuated
due to absorption by e$^+$e$^-$ pair creation with the photons of the Extragalactic Background Light (EBL).
The strength of the absorption depends on the distance (redshift) of the object, the energy of the VHE gamma-rays and
the SED of the EBL.
Observations in the VHE band therefore provide an indirect measurement of the
spectral energy density (SED) of the EBL \citep[][and references therein]{1992ApJ...390L..49S,1999APh....11...93P}.
In order to disentangle the effect of the EBL from
the intrinsic spectral shape of the objects,
observations of AGN over a wide range of redshifts and over a wide energy range are important.
The maximum of the absorption is given by the relation $1.24\,E_\gamma [\mathrm{TeV}] = \lambda [\mathrm{\mu m}]$ between
the energy of the gamma-ray and the wavelength of the EBL photon.
Spectral measurements beyond 20\,TeV would allow to constrain the mid- to far-infrared region of the EBL.

\section{H.E.S.S. observations and analysis}
\begin{figure}[ht]
\epsfig{file=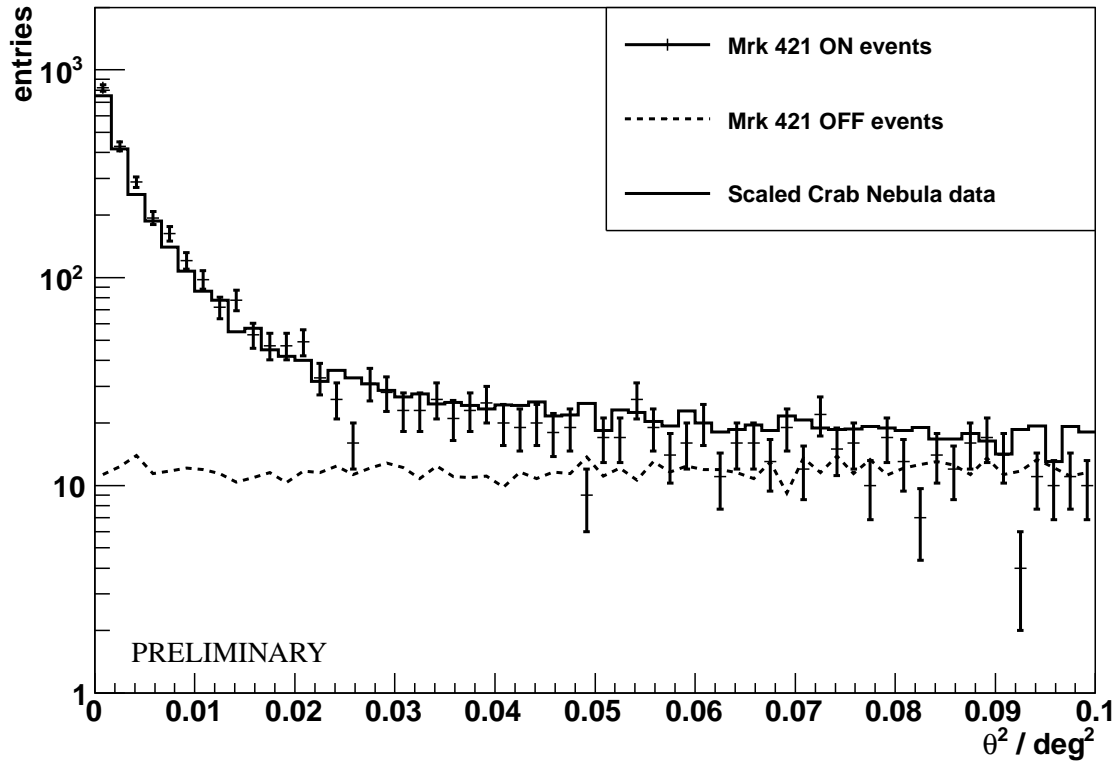,width=\columnwidth}
\parbox{\textwidth}{\vspace{-4cm}\hspace{2.0cm}PRELIMINARY}
\caption{\label{theta2}Distribution of reconstructed squared angular distance $\theta^2$. The observed excess in the
on-source region of Mrk\,421 is compatible with the point-spread function of the instrument. The corresponding distribution
from the Crab Nebula appears to be slightly flatter.}
\end{figure}

\begin{figure}[ht]
\epsfig{file=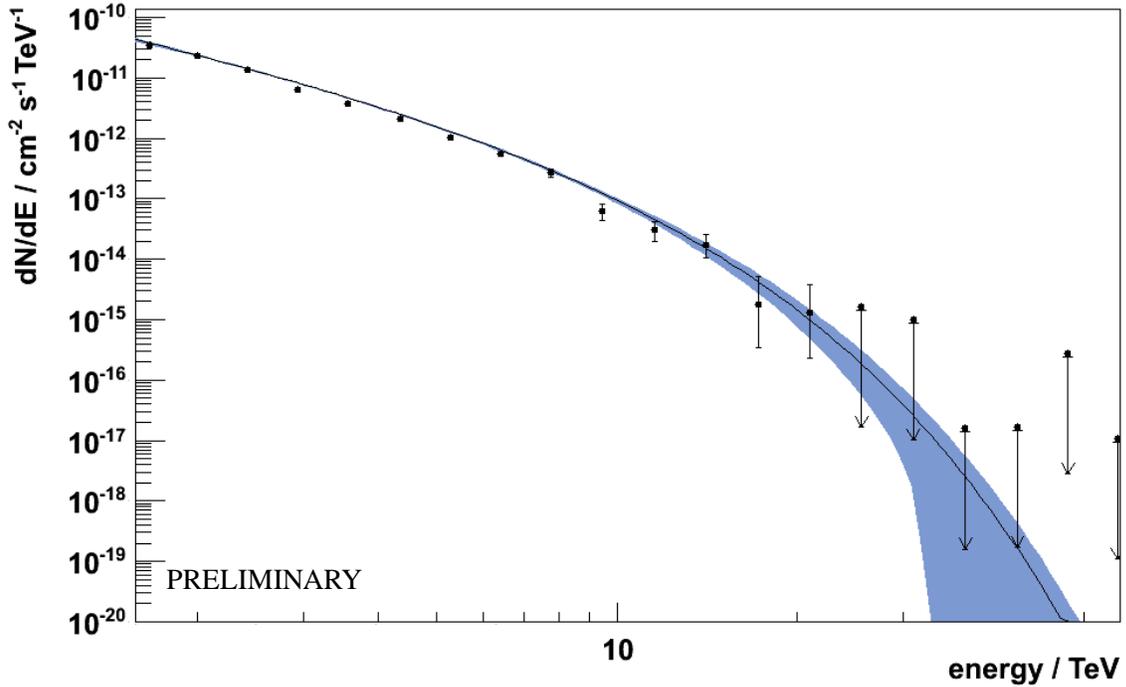,width=\columnwidth}
\parbox{\textwidth}{\vspace{-4cm}\hspace{2.2cm}PRELIMINARY}
\caption{\label{spectrum}Reconstructed H.E.S.S. spectrum, time-averaged for the whole flaring period in February
2010.}
\end{figure}

\begin{figure}[ht]
\epsfig{file=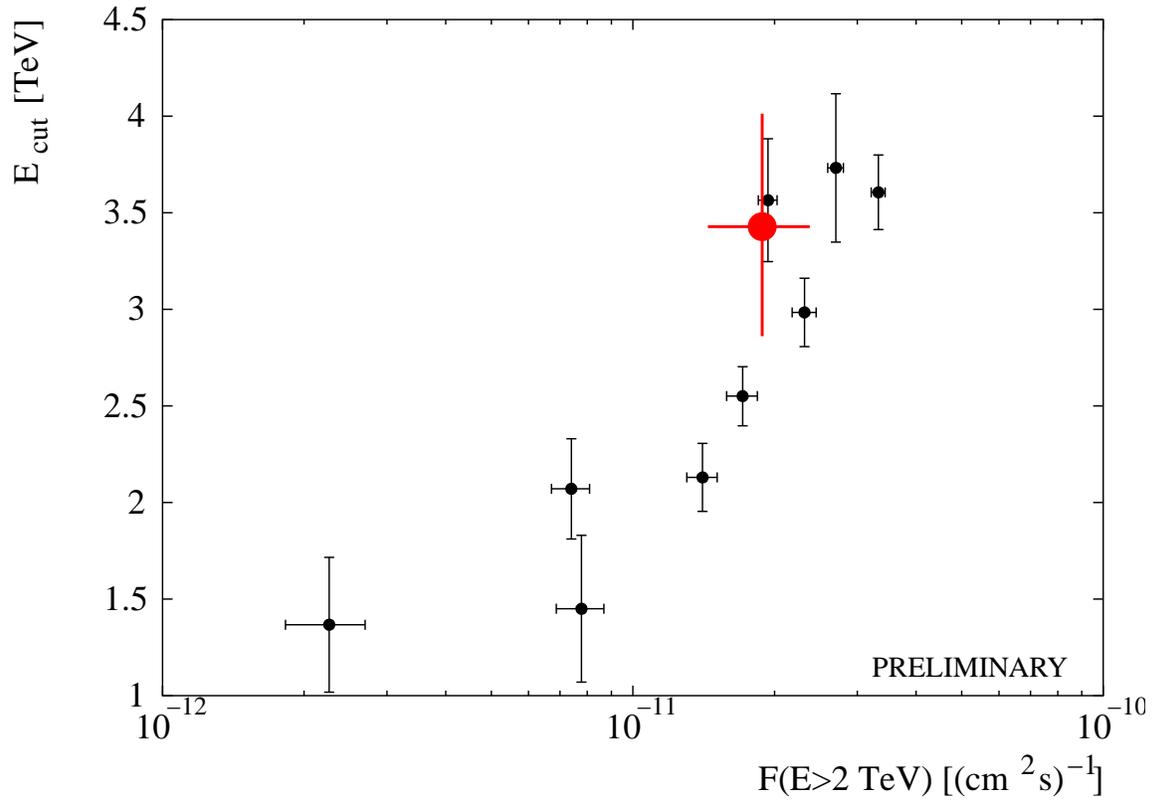,width=\columnwidth}
\parbox{\textwidth}{\vspace{-4.5cm}\hspace{11.5cm}PRELIMINARY}
\caption{\label{comparespectrum}Correlation between the observed flux state and the measured cutoff energy for different
flux states in 2004 (small full circles) and the time-averaged flux and cutoff energy of the present 2010 data set (large full circle). This
figure was adapted from \citet{Aharonian:2005ib}.}
\end{figure}

\begin{figure}[ht]
\epsfig{file=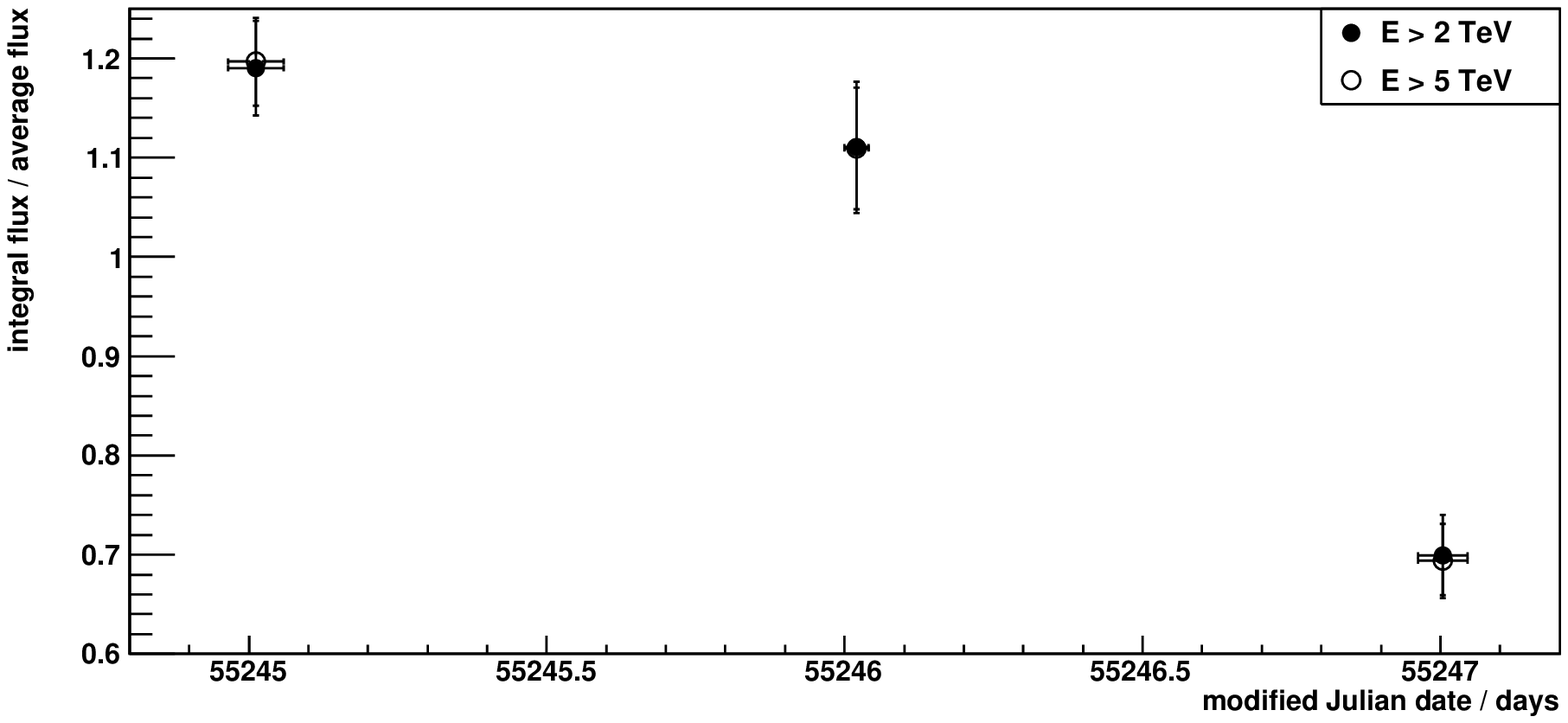,width=\columnwidth}
\epsfig{file=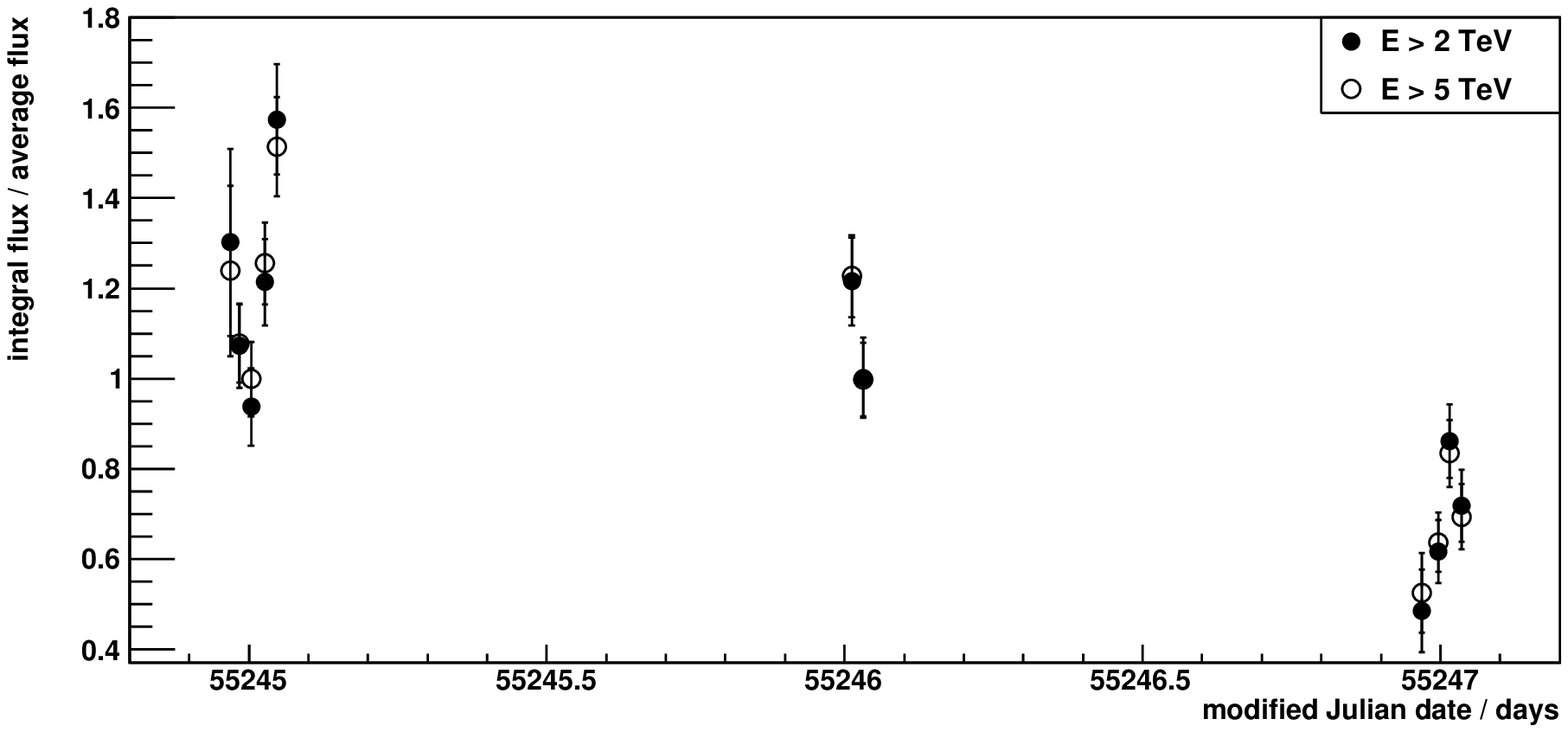,width=\columnwidth}
\parbox{\textwidth}{\vspace{-3.3cm}\hspace{2.2cm}PRELIMINARY}
\caption{\label{lc}Normalized night-by-night (top) and run-by-run (bottom)
lightcurves of Mrk\,421 during the H.E.S.S. observations of the February
2010 flare. Two thresholds for the calculation of the integral flux levels were chosen: 2\,TeV and 5\,TeV
(beyond the measured cut-off energy).
The level of variability is identical in both energy bands.
}
\end{figure}

\paragraph{Data}
The H.E.S.S. observations from February 2010 presented here
were triggered by the fast communication of a VERITAS high-state detection
\citep{2010ATel.2443....1C}.
A total of 6.5h of observation time were gathered by the H.E.S.S. telescopes from 17th to 20th of February
(MJD 55244.96 to MJD 55246.96). The data were taken in 11 runs (of 28 minutes each)
in the wobble observation mode, with an alternating source offset relative to the
camera center of 0.5$^\circ$.
With
62.4$^\circ$
the average zenith angle of this data set is very large,
resulting in a higher energy threshold as compared to observations of southern sources,
or observations of Mrk\,421 in the northern hemisphere by MAGIC or VERITAS,
and also leads to an enhanced effective area of the H.E.S.S. observations above 10\,TeV.
A standard run quality selection (based on trigger rate, and camera performance level)
followed by an additional quality cut on the muon efficiency corrected trigger
rate \citep{fernandes:2009a} were applied,
leaving 5.4 hours of live-time corrected high-quality data.
\paragraph{Signal}
These data were analysed using a likelihood reconstruction method \citep[\emph{model++},][]{2009APh....32..231D}.
Standard cuts on the squared angular distance ($\theta^2<0.01$), the minimum total charge in the camera ($>$60\,p.e.),
and the nominal distance of the camera images ($<$\,2.5\,deg).
The results from the independent standard online-analysis yielded compatible results.
The background was estimated from 11 off-regions \citep{aha:multioff} at the same distance to the camera center as the
on-region (centered around Mrk\,421).
A total of 2112 excess events (2188 on-events and 838/11 off-events) were detected in this dataset,
yielding a significance of 86.5 standard deviations.
The distribution of the square of the reconstructed angular distance $\theta^2$ is shown in Figure~\ref{theta2}
for the on- and off-regions. A clear signal at the position of Mrk\,421 is visible in the on-source distribution
as excess towards small values of $\theta^2$.
For comparison, the on-source distribution of the Crab Nebula (scaled to the Mrk\,421 distribution) is also shown.
The Mrk\,421 distribution is compatible to the point-spread function of the H.E.S.S. telescopes. The distribution
from the Crab Nebula appears to be slightly flatter.
\paragraph{Spectrum}
The time-averaged energy spectrum of the February data was calculated using a forward folding method under the assumption of
a powerlaw spectrum with exponential cut-off.
The resulting spectral shape is shown as shaded (1\,$\sigma$-contour) area in Figure~\ref{spectrum} along
with spectral points calculated on the basis of the found shape.
At least 3$\sigma$ are required for each point. The last significant bin extends beyond 20 TeV
(highest energy observed from this source), making these observations relevant for studies of the density
of the mid-infrared region of the EBL spectrum.
The spectral shape can be parameterized by $dN/dE = \phi_0 (E/E_0)^{-\Gamma} \mathrm{exp}(-E/E_{cut})$,
with a flux normalization $\phi_0 = (1.96\,\pm\,0.32)\,10^{-11}$\,cm$^{-2}$s$^{-1}$TeV$^{-1}$,
a spectral index $\Gamma = 2.05\,\pm\,0.22$, a cutoff energy $E_{cut} = (3.4\,\pm\,0.6)$\,TeV,
and a decorrelation energy $E_0 = 2.739$\,TeV.
The equivalent $\chi^2$-value from the forward-folding likelihood method
is 16.8 with 17 degrees of freedom.
Systematic errors are of the order 0.9 on the cutoff energy and 20\,\% on the flux.
The 2010 flare data yield a spectral shape that is compatible with previous
H.E.S.S. Results \citep{Aharonian:2005ib}.
As can be seen in Figure~\ref{comparespectrum}, the measured time-averaged flux and cutoff energy of the 2010
data are compatible with the previous H.E.S.S. observations in 2004.
These results confirm that the cutoff energy is significantly lower than in Mrk 501
($\approx$6.2 TeV). As concluded in \citet{Aharonian:2005ib}, given that both
objects are located at similar redshifts (0.031 / 0.033), this difference
cannot be explained with absorption by the EBL and therefore,
implies an intrinsic origin of the Mrk 421 energy cutoff.
The apparent stability of the cutoff-energy over the epochs 2004 and 2010
also implies a corresponding stability of the internal conditions of the central engine of Mrk421 \citep[also see][]{da09}.
\paragraph{Variability}
Integral fluxes above 2\,TeV and 5\,TeV were calculated for each night and on a run-by-run basis.
The results are shown in Figure~\ref{lc} and are summarized in Tables~\ref{tlc_night}~and~\ref{tlc_run}.
While clear variability is seen on a night-by-night basis, the significance of the run-by-run variability
(time-scale 0.5\,h) is low. For illustration, the strongest evidence for intra-night variability is seen in
the first observation night with a $\chi^2$-test p-value of $8\times10^{-4}$.
Further tests on shorter time-scales yielded no significant variability in these data.
The cosmic ray flux is consistent (on all tested time-scales) with a constant background
throughout the whole observation period.
\begin{table}[t]
\centering
\begin{tabular}{llll}
\hline
MJD start & MJD end & F(E$>$2\,TeV) & F(E$>$5\,TeV) \\\hline
55244.9637 &55245.0571  & 2.40$\pm$0.10 & 0.307$\pm$0.011\\
55246.0004 &55246.0411  & 2.24$\pm$0.13 & 0.284$\pm$0.016\\
55246.963  &55247.0469  & 1.41$\pm$0.08 & 0.178$\pm$0.010\\\hline
\end{tabular}
\caption{\label{tlc_night}Night-by-night integral flux values above 2\,TeV and 5\,TeV.}
\end{table}
\begin{table}[t]
\centering
\begin{tabular}{llll}
\hline
MJD start & MJD end & F(E$>$2\,TeV) & F(E$>$5\,TeV) \\\hline
55244.9637 &55244.9700  & 2.63     $\pm$0.42 &0.317   $\pm$0.048\\
55244.9741 &55244.9936  & 2.17     $\pm$0.19 &0.275   $\pm$0.022\\
55244.9950 &55245.0145  & 1.90     $\pm$0.17 &0.255   $\pm$0.021\\
55245.0166 &55245.0362  & 2.45     $\pm$0.19 &0.321   $\pm$0.023\\
55245.0376 &55245.0571  & 3.18     $\pm$0.25 &0.387   $\pm$0.028\\
55246.0004 &55246.0200  & 2.46     $\pm$0.20 &0.314   $\pm$0.023\\
55246.0215 &55246.0411  & 2.03     $\pm$0.18 &0.255   $\pm$0.021\\
55246.9630 &55246.9725  & 0.98     $\pm$0.19 &0.134   $\pm$0.022\\
55246.9851 &55247.0046  & 1.25     $\pm$0.14 &0.163   $\pm$0.017\\
55247.0062 &55247.0258  & 1.74     $\pm$0.16 &0.213   $\pm$0.019\\
55247.0273 &55247.0469  & 1.45     $\pm$0.16 &0.177   $\pm$0.019\\\hline
\end{tabular}
\caption{\label{tlc_run}Run-by-run integral flux values above 2\,TeV and 5\,TeV.}
\end{table}
The present February 2010 H.E.S.S. data show a flux level varying from 1.4 to 4.8 Crab flux units.
Previous observations of flaring episodes seen by other experiments have measured higher flux levels
\citep[see e.g.][and references therein]{2010A&A...524A..48T}.

\section{Summary}
A flaring episode of the BL Lac object Mrk421 was observed for 5.4h from the southern hemisphere at very large zenith angles by the H.E.S.S. Telescope system in February 2010. The observations show a similar level of activity as previously observed by H.E.S.S. The measured spectral shape is compatible to the previous results, implying a stability of the central engine over the epochs. The energy range of these observations extends up to 20 TeV primary gamma-ray energy, the highest energy observed from Mrk 421 so far. The spectral reach of these data make them relevant for studies of the extragalactic background light.

\end{document}